%% file: lnv-pipiee.tex
\documentclass[11pt]{article}
\usepackage[dvips]{graphicx}
\usepackage{epsfig}
\usepackage{amssymb}
\usepackage{color}
\usepackage{amsmath}
\usepackage{nicefrac}
\usepackage[left]{lineno}
\usepackage{hyperref}
\usepackage{adjustbox}
\usepackage{xspace}
\usepackage{multirow}
\usepackage{xcolor}

\definecolor{blu}{rgb}{0.,0.,1.}
\definecolor{red}{rgb}{1.,0.,0.}
\definecolor{burgundy}{rgb}{0.5, 0.0, 0.13}
\definecolor{crimsonred}{rgb}{0.6, 0.0, 0.0}
\definecolor{persianblue}{rgb}{0.11, 0.22, 0.73}
\definecolor{forestgreen}{rgb}{0.13,0.35,0.13}

\def\geant {\mbox{\textsc{Geant4}}\xspace}

\hypersetup{colorlinks, citecolor=crimsonred, linkcolor=persianblue, urlcolor=crimsonred}

\begin{document}
\centerline{\LARGE EUROPEAN ORGANIZATION FOR NUCLEAR RESEARCH}

%
\vspace{10mm} {\flushright{
CERN-EP-2022-018 \\
1 Feb 2022\\
\vspace{4mm}
Revised version:\\8 May 2022\\
}}
\vspace{-30mm}

%

%
\vspace{40mm}

\begin{center}
\boldmath
{\bf {\Large\boldmath{Searches for lepton number violating $K^+\to\pi^-(\pi^0)e^+e^+$ decays}}}
\unboldmath
\end{center}
\vspace{4mm}
\begin{center}
{\Large The NA62 Collaboration}\\
\end{center}

\begin{abstract}
Searches for lepton number violating $K^+\to\pi^-e^+e^+$ and $K^+\to\pi^-\pi^0e^+e^+$ decays have been performed using the complete dataset collected by the NA62 experiment at CERN in 2016--2018. Upper limits of $5.3\times 10^{-11}$ and $8.5\times 10^{-10}$ are obtained on the decay branching fractions at 90\% confidence level. The former result improves by a factor of four over the previous best limit, while the latter result represents the first limit on the $K^+\to\pi^-\pi^0e^+e^+$ decay rate.
\end{abstract}

\begin{center}
{\it Accepted for publication in Physics Letters B}
\end{center}

\newpage
\input{authors}
\newpage


\section*{Introduction}

In the Standard Model (SM), neutrinos are strictly massless due to the absence of right-handed chiral states. The discovery of neutrino oscillations has conclusively demonstrated the non-zero neutrino mass, making it possible, in principle, to distinguish experimentally between the Dirac and Majorana neutrino. In the minimal Type-I seesaw model~\cite{mo80}, the neutrino is a Majorana fermion with a mass term that violates lepton number by two units. Strong evidence for the Majorana nature of the neutrino would be provided by observation of lepton number violating (LNV) processes, such as charged kaon decays~\cite{li00}. The existing experimental limits on $K^+\to\pi^-\ell^+_1\ell^+_2$ decays lead to stringent constraints on active-sterile mixing angles between Majorana neutrinos. Below the kaon mass, these constraints are competitive with those obtained from neutrinoless double beta decays~\cite{atre05,atre09,ab18}.

The NA62 experiment at CERN collected a large dataset of $K^+$ decays into lepton pairs in 2016--2018, using dedicated trigger lines. Part of this dataset has been analysed to establish upper limits on the rates of lepton number and flavour violating decays $K^+\to\pi^-\ell^+\ell^+$ ($\ell=e,\mu$)~\cite{co19}, and the full dataset has been analysed to obtain limits on $K^+\to\pi^\pm\mu^\mp e^+$ and $\pi^0\to\mu^-e^+$ decays~\cite{al21}, improving by up to an order of magnitude on earlier results. Searches for the $K^+\to\pi^-e^+e^+$ and $K^+\to\pi^-\pi^0e^+e^+$ decays based on the complete NA62 dataset collected in 2016--2018 are reported here, representing the first search for the latter process.

\section{Beam, detector and data sample}
\label{sec:detector}

\begin{figure}[t]
\begin{center}
\vspace{-2mm}
\resizebox{\textwidth}{!}{\includegraphics{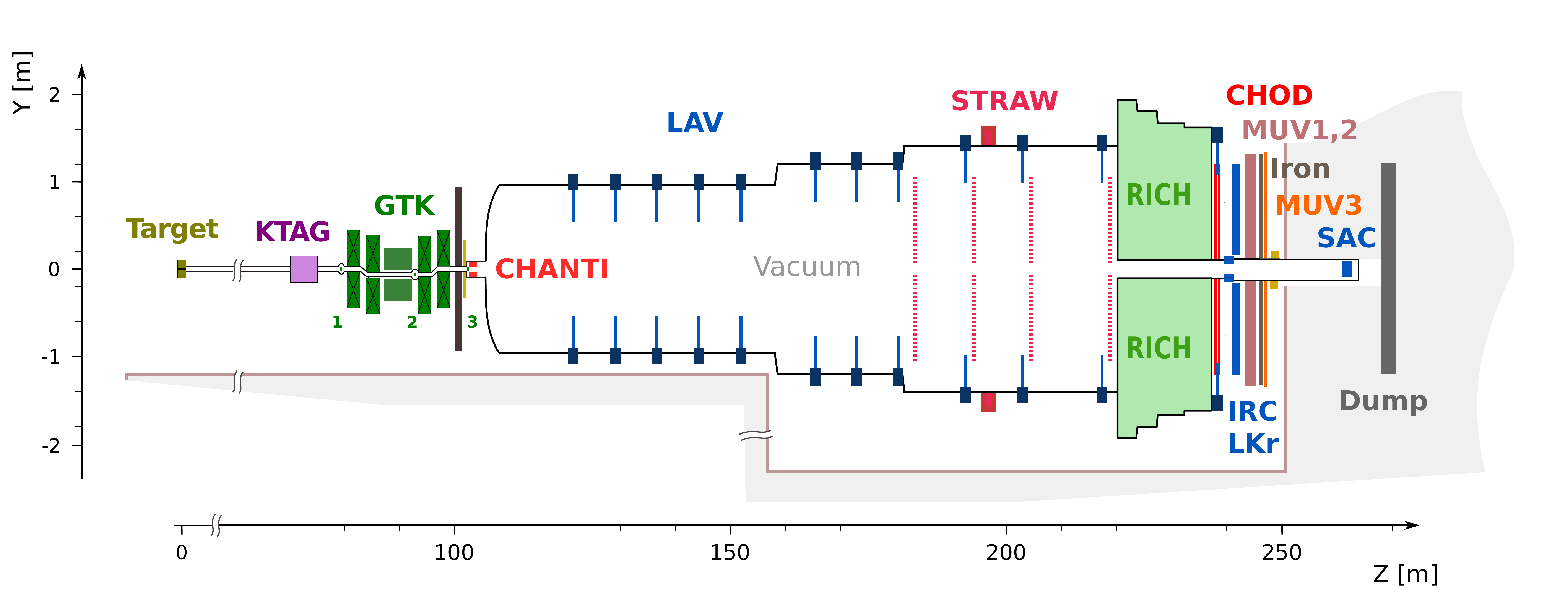}}
\put(-343,67){\scriptsize\color{burgundy}\rotatebox{90}{\textbf{\textsf{SCR}}}}
\put(-325,121){\scriptsize\color{burgundy}\rotatebox{90}{\textbf{\textsf{COL}}}}
\put(-172,46){\scriptsize\color{red}{\textbf{\textsf{M}}}}
\end{center}
\vspace{-17.5mm}
\caption{Schematic side view of the NA62 beamline and detector.}
\label{fig:detector}
\end{figure}

The layout of the NA62 beamline and detector~\cite{na62-detector} is shown schematically in Fig.~\ref{fig:detector}. An unseparated secondary beam of $\pi^+$ (70\%), protons (23\%) and $K^+$ (6\%) is created by directing 400~GeV/$c$ protons extracted from the CERN SPS onto a beryllium target in spills of 3~s effective duration. The central beam momentum is 75~GeV/$c$, with a momentum spread of 1\% (rms).

Beam kaons are tagged with a time resolution of 70~ps by a differential Cherenkov counter (KTAG), which uses nitrogen gas at 1.75~bar pressure contained in a 5~m long vessel as radiator. Beam particle positions, momenta and times (to better than 100~ps resolution) are measured by a silicon pixel spectrometer consisting of three stations (GTK1,2,3) and four dipole magnets. A muon scraper (SCR) is installed between GTK1 and GTK2. A 1.2~m thick steel collimator (COL) with a $76\times40$~mm$^2$ central aperture and $1.7\times1.8$~m$^2$ outer dimensions is placed upstream of GTK3 to absorb hadrons from upstream $K^+$ decays; a variable aperture collimator of \mbox{$0.15\times0.15$~m$^2$} outer dimensions was used up to early 2018. Inelastic interactions of beam particles in GTK3 are detected by an array of scintillator hodoscopes (CHANTI). The beam is delivered into a vacuum tank evacuated to a pressure of $10^{-6}$~mbar, which contains a 75~m long fiducial volume (FV) starting 2.6~m downstream of GTK3. The beam angular spread at the FV entrance is 0.11~mrad (rms) in both horizontal and vertical planes. Downstream of the FV, undecayed beam particles continue their path in vacuum.

Momenta of charged particles produced in $K^+$ decays in the FV are measured by a magnetic spectrometer (STRAW) located in the vacuum tank downstream of the FV. The spectrometer consists of four tracking chambers made of straw tubes, and a dipole magnet (M) located between the second and third chambers that provides a horizontal momentum kick of 270~MeV/$c$. The momentum resolution is $\sigma_p/p = (0.30\oplus 0.005 p)\%$, with the momentum $p$ expressed in GeV/$c$.

%
A ring-imaging Cherenkov detector (RICH) consisting of a 17.5~m long vessel filled with neon at atmospheric pressure (with a Cherenkov threshold of 12.5~GeV/$c$ for pions) provides particle identification, charged particle time measurements with a typical resolution of 70~ps, and the trigger time.
The RICH optical system is optimised to collect light emitted by positively charged particles, exploiting their deflection by the STRAW dipole magnet. Two scintillator hodoscopes (CHOD), which include a matrix of tiles and two planes of slabs arranged in four quadrants located downstream of the RICH, provide trigger signals and time measurements with 200~ps precision.


A $27X_0$ thick quasi-homogeneous liquid krypton (LKr) electromagnetic calorimeter is used for particle identification and photon detection. The calorimeter has an active volume of 7~m$^3$ segmented in the transverse direction into 13248 projective cells of
$2\times 2$~cm$^2$ size, and provides an energy resolution $\sigma_E/E=(4.8/\sqrt{E}\oplus11/E\oplus0.9)\%$, with $E$ expressed in GeV. To achieve hermetic acceptance for photons emitted in $K^+$ decays in the FV at angles up to 50~mrad from the beam axis, the LKr calorimeter is supplemented by annular lead glass detectors (LAV) installed in 12~positions inside and downstream of the vacuum tank, and two lead/scintillator sampling calorimeters (IRC, SAC) located close to the beam axis. An iron/scintillator sampling hadronic calorimeter formed of two modules (MUV1,2) and a muon detector consisting of 148~scintillator tiles located behind an 80~cm thick iron wall (MUV3) are used for particle identification.

%
%

The data sample analysed is obtained from $0.89\times 10^6$ SPS spills recorded in 2016--2018, with the typical beam intensity increasing over time from \mbox{$1.3\times 10^{12}$} to \mbox{$2.2\times 10^{12}$} protons per spill. The latter value corresponds to a 500~MHz mean instantaneous beam particle rate at the FV entrance, and a 3.7~MHz mean $K^+$ decay rate in the FV. Multi-track (MT) and electron multi-track ($e$MT) trigger chains downscaled typically by factors of 100 and 8, respectively, are used for the analysis. The low-level (L0) trigger~\cite{am19} for both chains is based on RICH signal multiplicity and coincidence of signals in two opposite CHOD quadrants. The $e$MT chain additionally involves a requirement of at least 20~GeV energy deposit in the LKr calorimeter. The high-level software (L1) trigger used for both chains involves beam $K^+$ identification by the KTAG and reconstruction of a negatively charged STRAW track. For signal-like samples (which are characterised by LKr energy deposit well above 20~GeV),
the measured inefficiencies of the CHOD (STRAW) trigger conditions are typically at the
1\%~(5\%) level, while those of the RICH, KTAG and LKr are ${\cal O}(10^{-3})$.

Monte Carlo (MC) simulations of particle interactions with the detector and its response are performed with a software package based on the~\geant toolkit~\cite{geant4}, and include pileup (i.e. coincidences of multiple events in time) and the full trigger chain.


\vspace{-1mm}
\section{Event selection}
\label{sec:selection}
\vspace{-1mm}

The rates of the possible signal decays $K^+\to\pi^-(\pi^0)e^+e^+$ are measured with respect to the rate of the normalisation decay $K^+\to\pi^+e^+e^-$. This approach leads to significant cancellation of the effects of detector inefficiencies, trigger inefficiencies and pileup. The signal and normalisation processes are collected concurrently using the MT and $e$MT trigger chains described above. The main selection criteria are listed below.
\begin{itemize}
\item Three-track vertices are reconstructed by backward extrapolation of STRAW tracks into the FV, taking into account the measured residual magnetic field in the vacuum tank, and selecting triplets of tracks consistent with originating from the same point. Exactly one such vertex should be present in the event. The total charge of the tracks forming the vertex should be $q=1$, and the vertex longitudinal position ($z_{\rm vtx}$) should be within the FV. The momentum of each track forming the vertex should be in the range 6--44~GeV/$c$, and its trajectory through the STRAW chambers, and its extrapolation to the CHOD and LKr calorimeter, should be within the respective geometrical acceptances. Each pair of tracks should be separated by at least 15~mm in each STRAW chamber plane to suppress photon conversions, and by at least 200~mm in the LKr front plane to reduce shower overlap effects.
\item Track times are initially defined using the CHOD information. The vertex CHOD time is evaluated as the average of track CHOD times. The RICH signal pattern within 3~ns of the vertex CHOD time is used to compute the likelihoods of mass hypotheses for each track and evaluate track RICH times. Track and vertex time estimates are then recomputed using the RICH information. Each track is required to be within 2.5~ns of the trigger time.
\item To suppress backgrounds from $K^+\to\pi^+\pi^0_D$ and $K^+\to\pi^0_D e^+\nu$ decays followed by the Dalitz decay $\pi^0_D\to\gamma e^+e^-$, which are characterised by emission of soft photons at large angles, no signals within 4~ns of the vertex time are allowed in the LAV detectors located downstream of the reconstructed vertex position. Since energetic photons emitted forward are already suppressed by the total momentum condition (see below), no photon veto requirements are applied in the LKr, IRC and SAC calorimeters. 
\item The ratio of energy deposited in the LKr calorimeter to the momentum measured by the spectrometer, $E/p$, identifies pion ($\pi^\pm$) and electron ($e^\pm$) candidates: $E/p<0.85$ for pions, and $0.9<E/p<1.1$ for electrons. The vertex should consist of a $\pi^\pm$ candidate and two $e^\pm$ candidates.
\end{itemize}
The following condition is applied to select $K^+\to\pi^+e^+e^-$ and $K^+\to\pi^-e^+e^+$ decays.
\begin{itemize}
\item The total momentum of the three tracks, $p_{\pi ee}$, should satisfy $|p_{\pi ee}-p_{\rm beam}|<2~{\rm GeV}/c$, where $p_{\rm beam}$ is the central beam momentum. The total transverse momentum with respect to the beam axis should be $p^{\pi ee}_T<30~{\rm MeV}/c$. The quantity $p_{\rm beam}$ and the beam axis direction are monitored throughout the data taking period using fully reconstructed $K^+\to\pi^+\pi^+\pi^-$ decays.
\end{itemize}
The following conditions are applied to select $K^+\to\pi^+e^+e^-$ decays.
\begin{itemize}
\item The reconstructed $e^+e^-$ mass should be $m_{ee}>140~{\rm MeV}/c^2$ to suppress backgrounds from the $K^+\to\pi^+\pi^0$ decay followed by $\pi^0_D\to e^+e^-\gamma$, $\pi^0_{DD}\to e^+e^-e^+e^-$ and $\pi^0\to e^+e^-$ decays. This leads to a 27\% fractional loss in acceptance.
\item The signal region of reconstructed $\pi^+e^+e^-$ mass, $m_{\pi ee}$, is defined as 470--505~MeV/$c^2$, which accounts for the mass resolution of 1.7~MeV/$c^2$ and the radiative mass tail. The lower mass region, $m_{\pi ee}<470~{\rm MeV}/c^2$, is used for validation of the background estimates.
\end{itemize}
The following conditions are applied to select $K^+\to\pi^-e^+e^+$ decays.
\begin{itemize}
\item RICH-based $e^+$ identification suppresses the otherwise dominant backgrounds from $K^+\to\pi^+\pi^0_D$ and $K^+\to\pi^+e^+e^-$ decays with double $\pi^+\to e^+$ and $e^-\to\pi^-$ misidentification. The identification condition is based on the likelihoods of $e^+$ and $\pi^+$ mass hypotheses evaluated using the RICH signal pattern. Additionally, the angles between track pairs in the RICH are required to exceed 4~mrad to reduce overlaps of Cherenkov light-cones, which causes a fractional reduction of 7\% in the signal acceptance.
\item The signal region of reconstructed $\pi^-e^+e^+$ mass, $m_{\pi ee}$, is defined as 488.6--498.8~MeV/$c^2$, corresponding to six times the mass resolution of 1.7~MeV/$c^2$. The mass region 470--505~MeV/$c^2$, which includes the signal region, is kept masked until the validation of the background estimate. The lower and upper mass regions used for validation of the background estimate are defined as $m_{\pi ee}<470~{\rm MeV}/c^2$ and $505~{\rm MeV}/c^2<m_{\pi ee}<600~{\rm MeV}/c^2$, respectively.
\end{itemize}
The following conditions are applied to select $K^+\to\pi^-\pi^0e^+e^+$ decays.
\begin{itemize}
\item The $\pi^0$ is reconstructed by its prompt $\pi^0\to\gamma\gamma$ decay. Exactly two photon candidates are required, defined as reconstructed LKr energy deposit clusters with energy above 2~GeV, within 5~ns of the vertex time, separated by at least 150~mm from each other and from each track impact point in the nominal LKr transverse plane.
\item The longitudinal coordinate of the neutral vertex is defined
assuming that the two photons are emitted in a $\pi^0\to\gamma\gamma$ decay: $z_{\rm N} = z_{\rm LKr} - D_{12}\sqrt{E_1E_2}/m_{\pi 0}$.
Here $D_{12}$ is the distance between the two clusters in the LKr transverse plane (with a $z$ coordinate $z_{\rm LKr}$), $E_{1,2}$ are the photon candidate energies, and $m_{\pi 0}$ is the nominal $\pi^0$ mass~\cite{pdg}.
\item Consistency of the three-track and neutral vertices is required: $|z_{\rm vtx}-z_{\rm N}|<8$~m. Vertex position resolutions evaluated with simulations are $\delta z_{\rm vtx}=0.25$~m and $\delta z_{\rm N}=1.8$~m.
\item Photon momenta are computed using cluster energies and positions in the LKr transverse plane, assuming emission at the three-track vertex. The $\pi^0$ momentum is then evaluated as the sum of photon momenta.
%
\item The total final state momentum, $p_{\pi \pi e e}$, should be consistent with the central beam momentum: $|p_{\pi\pi ee}-p_{\rm beam}|<3~{\rm GeV}/c$. The total transverse momentum with respect to the beam axis is required to be $p^{\pi\pi ee}_T<30~{\rm MeV}/c$.
\item The signal region of reconstructed $\pi^-\pi^0e^+e^+$ mass, $m_{\pi\pi ee}$, is defined as 484--504~MeV/$c^2$, and is kept masked until the validation of the background estimates. The mass resolution is 1.9~MeV/$c^2$, and a loose signal region definition is adopted due to the lack of background. The control region 400--600~MeV/$c^2$ (excluding the signal region) is used for validation of the background estimate.
\end{itemize}

%
%
\begin{figure}[tb]
\begin{center}
\resizebox{0.5\textwidth}{!}{\includegraphics{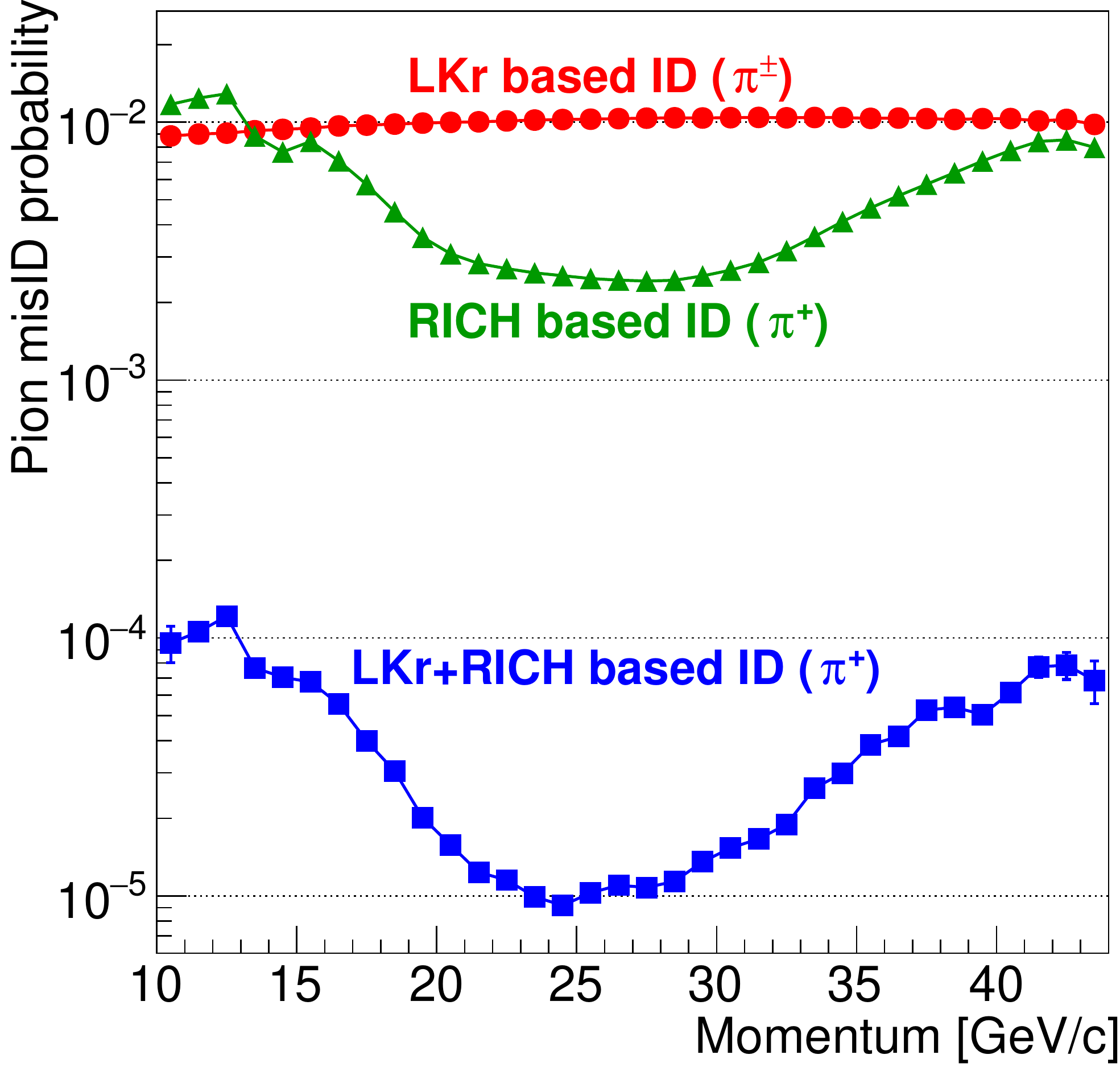}}
\resizebox{0.5\textwidth}{!}{\includegraphics{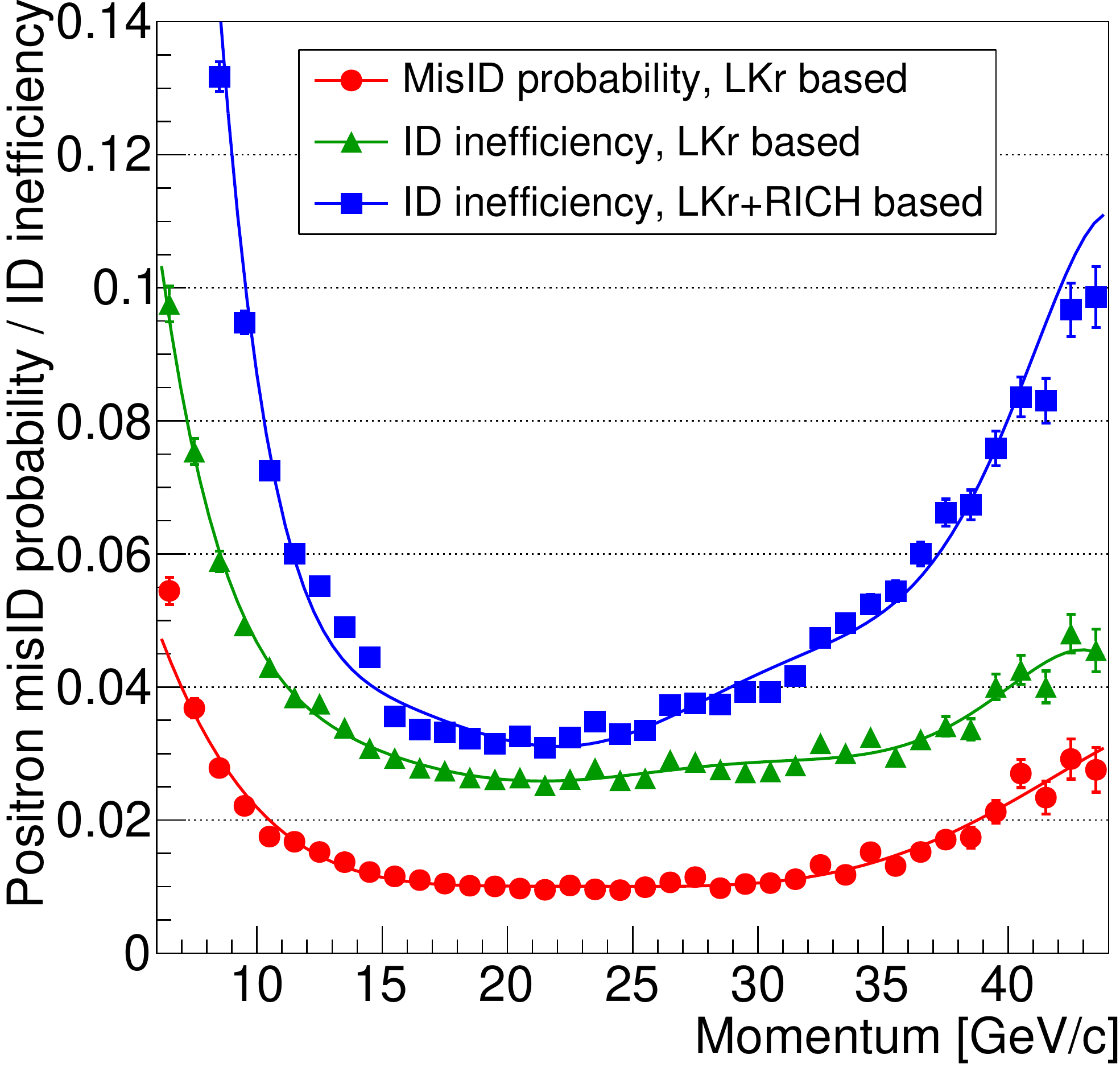}}
\end{center}
\vspace{-12mm}
\caption{Left: $\pi^\pm$ misidentification probability $P_{\pi e}$ for the LKr-based and RICH-based identification conditions, applied separately and together, measured in momentum bins. Right: $e^+$ misidentification probability $P_{e\pi}$ for the LKr-based $\pi^+$ identification condition, and $e^+$ identification inefficiency $1-\varepsilon_e$ for the LKr-based and RICH-based conditions, measured in momentum bins. Polynomial fits to the measured values used for the positron identification model are shown.}
\label{fig:pid}
\end{figure}


\section{Particle identification studies}
\label{sec:bkg}

Backgrounds to signal decays arise mainly from pion ($\pi^\pm$) misidentification as electron ($e^\pm$) and vice versa. As discussed in Section~\ref{sec:selection}, LKr-based identification is used for $\pi^\pm$ and $e^\pm$, and RICH-based identification is additionally employed for $e^+$ within the $K^+\to\pi^-e^+e^+$ selection. The accuracy of \geant-based particle identification simulation is limited: the quantity $E/p$ is sensitive to hadronic shower and cluster reconstruction simulation, while the RICH-based algorithm depends critically on gas pressure, light yield and mirror alignment calibrations. A dedicated data-driven model is used to simulate particle identification for the MC samples: the measured identification probabilities are applied as weights to MC events. This approach also improves the statistical precision on the estimated backgrounds.

Pion misidentification probability ($P_{\pi e}$) and identification efficiency ($\varepsilon_\pi$) are measured as functions of momentum in the range 10--44~GeV/$c$ using a pure $\pi^\pm$ sample obtained by kinematic selection of $K^+\to\pi^+\pi^+\pi^-$ decays. The lower bound of 10~GeV/$c$ represents the kinematic limit of $K^+\to\pi^+\pi^+\pi^-$ decays. The efficiency $\varepsilon_\pi$ varies from 98.2\% to 98.7\% as a function of momentum. The $P_{\pi e}$ measurements are shown in Fig.~\ref{fig:pid}~(left). The LKr-based $e^\pm$ identification leads to $P_{\pi e}\approx 10^{-2}$ with a weak momentum dependence, and $P_{\pi e}$ values are larger for $\pi^+$ than for $\pi^-$ by about $5\times 10^{-4}$ in absolute terms which is attributed to the larger $\pi^+$ charge exchange ($\pi^+n\to\pi^0p$) cross-section on Krypton nuclei. The RICH-based $e^+$ identification provides an additional $\pi^+$ rejection factor of up to $10^3$. The strongest suppression corresponds to the $\pi^+$ momentum range for which the RICH is optimised, while non-zero $P_{\pi e}$ values below the Cherenkov threshold are due to the presence of additional in-time tracks in the fully reconstructed $K^+\to\pi^+\pi^+\pi^-$ events. The model takes into account the dependence of the RICH-based $P_{\pi e}$ on the angle between the two $\pi^+$ tracks in the RICH, caused by ring overlaps. The model also accounts for the correlation between the measured track momentum $p$ and $E/p$.

Positron misidentification probability ($P_{e\pi}$) and identification efficiency ($\varepsilon_e$) are measured
in the full momentum range 6--44~GeV/$c$ used in the event selection using a positron sample obtained by kinematic selection of $K^+\to\pi^0e^+\nu$ decays. The background contamination from $K^+\to\pi^+\pi^0$ and $K^+\to\pi^0\mu^+\nu$ decays is estimated from simulations and subtracted: it varies from negligible at low momentum to 0.6\% at high momentum. The measurements are shown in Fig.~\ref{fig:pid}~(right): both $P_{e\pi}$ and $\varepsilon_e$ exhibit a significant momentum dependence. Small differences in the LKr calorimeter response for electrons and positrons have negligible effect on this analysis, and the $e^+$ measurements are also used to model $e^-$ identification for the MC samples.


\begin{figure}[p]
\begin{center}
\resizebox{0.5\textwidth}{!}{\includegraphics{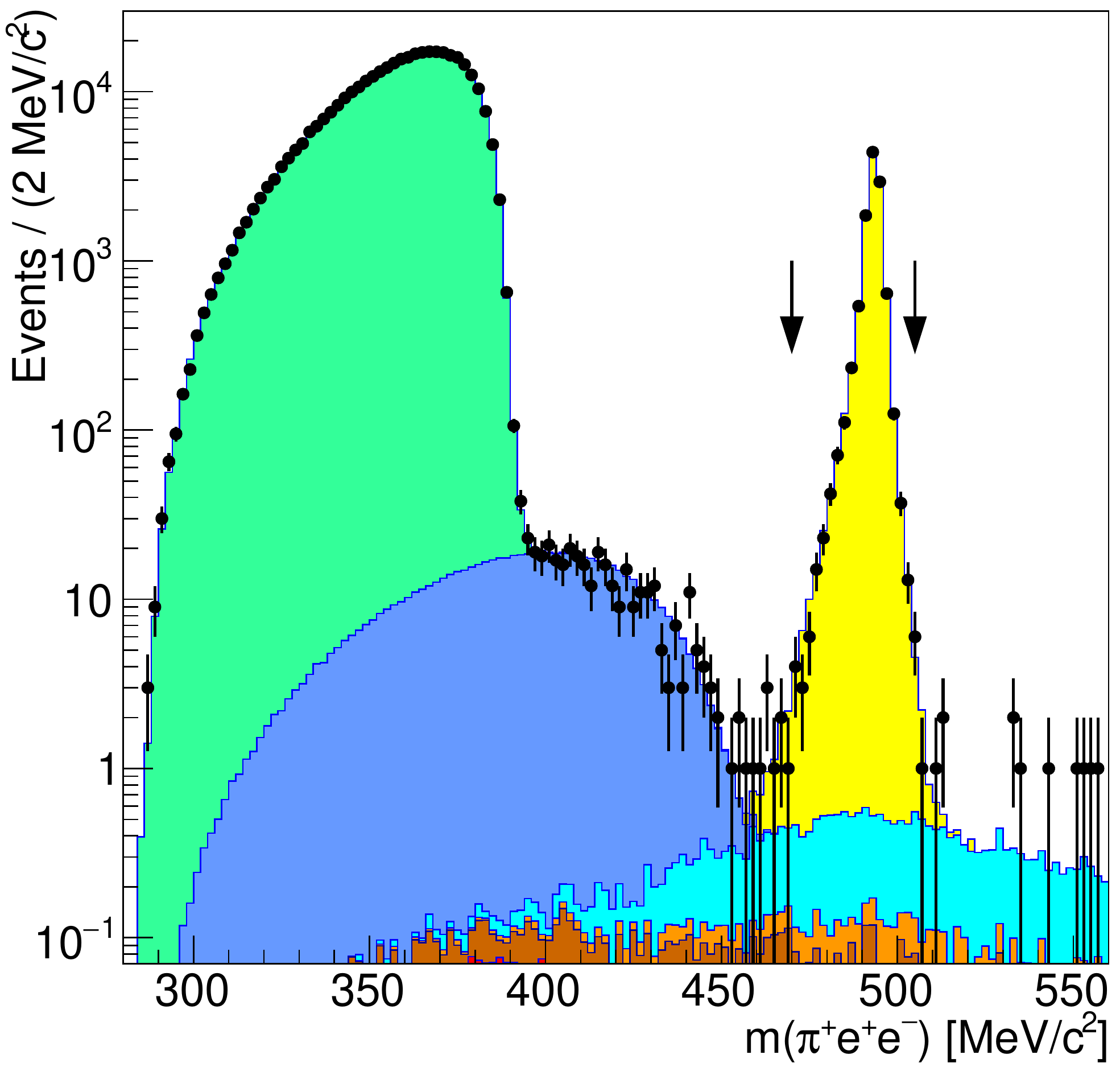}}%
\resizebox{0.5\textwidth}{!}{\includegraphics{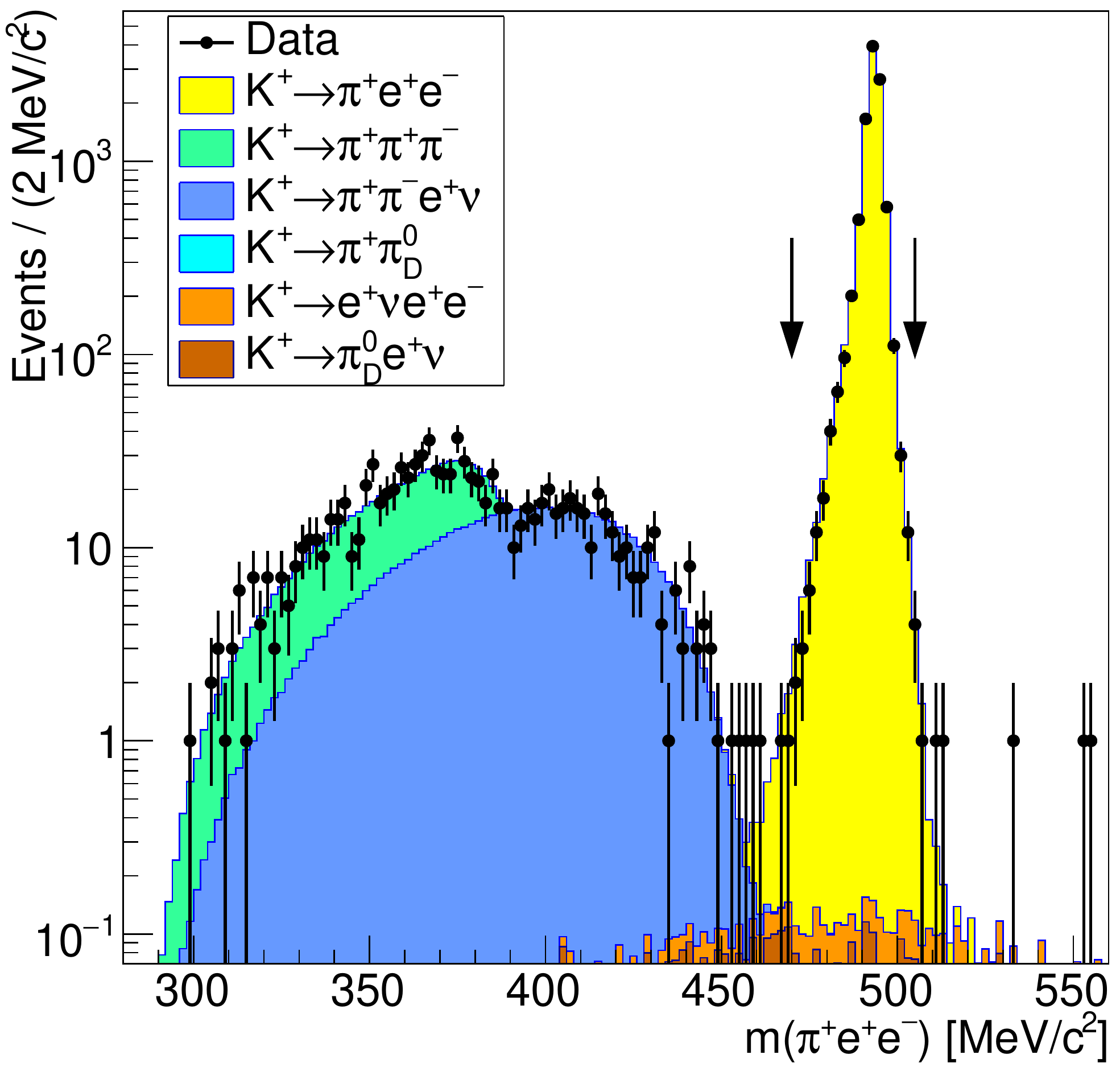}}
\end{center}
\vspace{-14mm}
\caption{Reconstructed $m_{\pi ee}$ spectra of the data, simulated signal and backgrounds obtained using the standard $K^+\to\pi^+e^+e^-$ selection (left) and the control $K^+\to\pi^+e^+e^-$ selection involving RICH-based $e^+$ identification (right). The signal $m_{\pi ee}$ region is indicated with arrows.}
\label{fig:piee}
\end{figure}

\begin{figure}[p]
\begin{center}
\resizebox{0.5\textwidth}{!}{\includegraphics{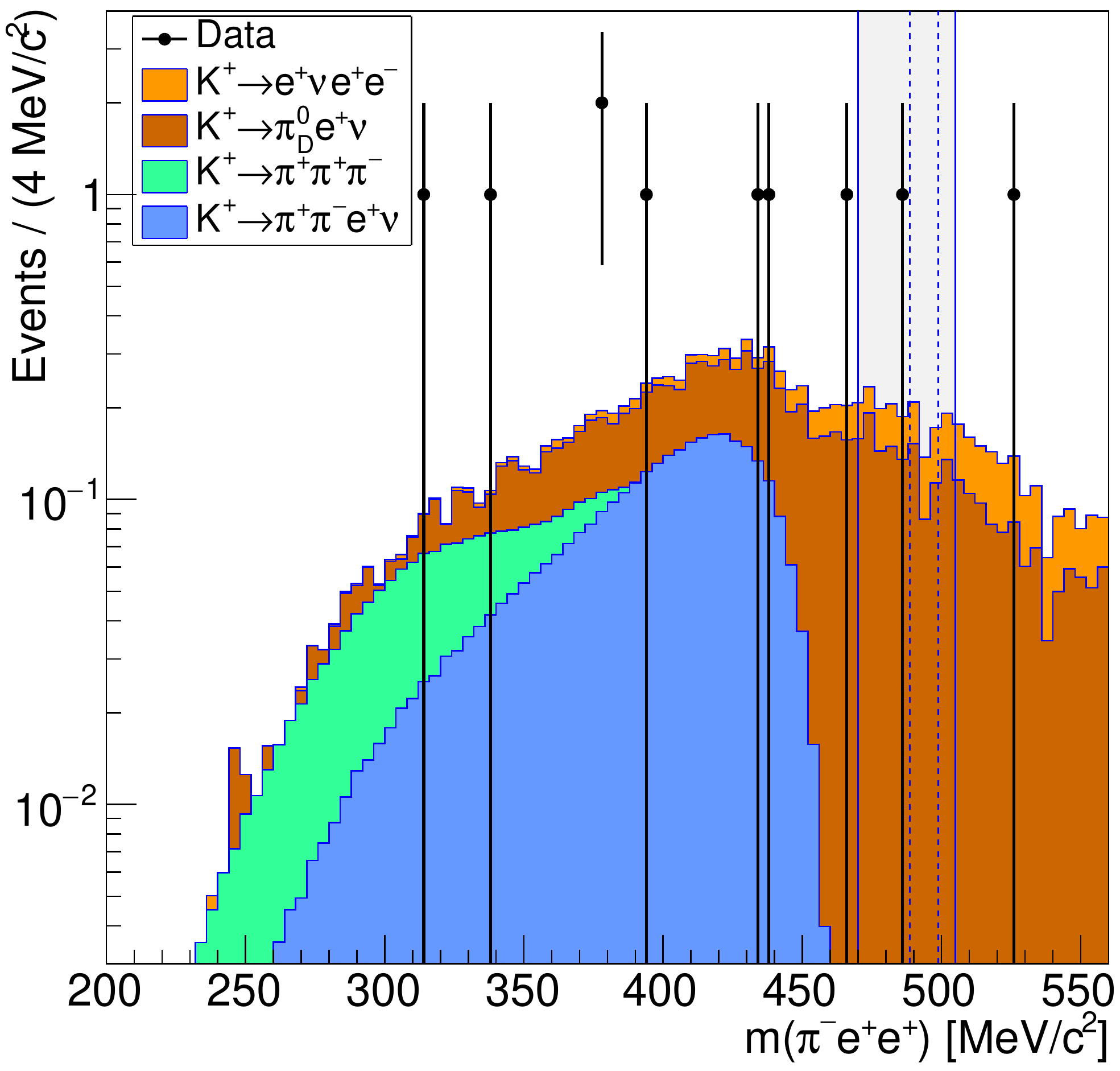}}%
\resizebox{0.5\textwidth}{!}{\includegraphics{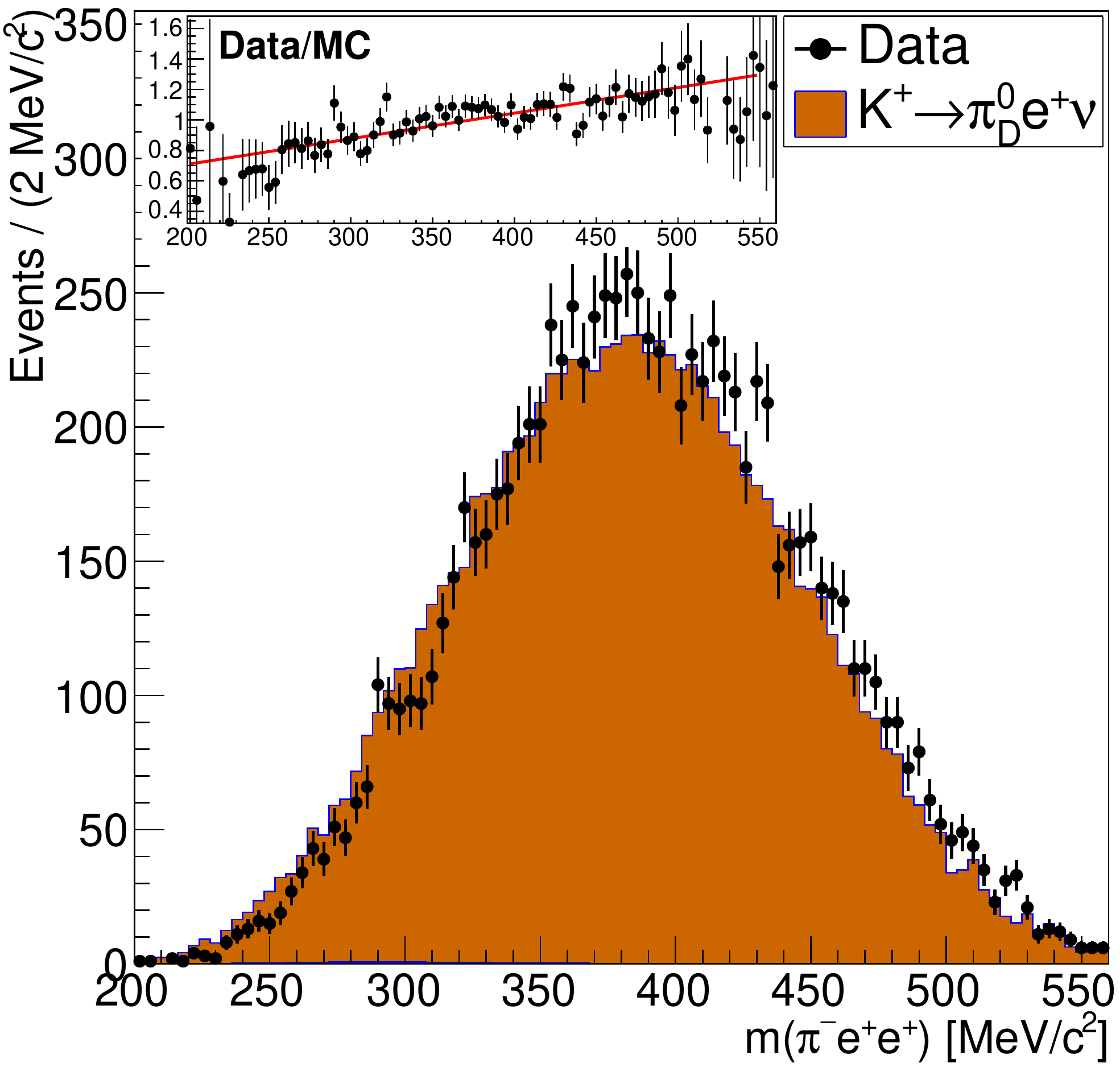}}
\end{center}
\vspace{-14mm}
\caption{Left: reconstructed $m_{\pi ee}$ spectra of the data and simulated backgrounds obtained using the $K^+\to\pi^-e^+e^+$ selection. The shaded vertical band indicates the $m_{\pi ee}$ region masked during the analysis, including the signal region bounded by dashed lines. Right: reconstructed $m_{\pi ee}$ spectra of the data and simulated backgrounds obtained using the control $K^+\to\pi^-e^+e^+$ selection with a missing momentum requirement. Inset: ratio of the data and simulated spectra, and a fit to the ratio with a linear function.}
\label{fig:piee-lnv}
\end{figure}


\boldmath
\section{Normalisation to the $K^+\to\pi^+e^+e^-$ decay}
\unboldmath
\label{sec:piee}

The $K^+\to\pi^+e^+e^-$ sample is used for normalisation of the estimated backgrounds, and validation of the $\pi^\pm$ misidentification modelling. In addition to the standard $K^+\to\pi^+e^+e^-$ selection relying on LKr-based particle identification, a control selection with RICH-based $e^+$ identification is considered for validation purposes. The reconstructed $m_{\pi ee}$ spectra of the data, simulated signal and backgrounds obtained with the two selections are displayed in Fig.~\ref{fig:piee}.

\newpage

The background in the signal $m_{\pi ee}$ region comes mainly from the $K^+\to\pi^+\pi^0_D$ decay with double misidentification ($\pi^+\to e^+$ and $e^+\to\pi^+$). Principal backgrounds in the lower $m_{\pi ee}$ region are due to the $K^+\to\pi^+\pi^+\pi^-$ decay with double $\pi^\pm\to e^\pm$ misidentification, and the $K^+\to\pi^+\pi^-e^+\nu$ decay with $\pi^-\to e^-$ misidentification. The former source is reduced by a factor ${\cal O}(10^3)$ by the RICH-based $e^+$ identification. To model the latter source, a constant $P_{\pi e}$ value is assumed for LKr-based identification for pion momenta below 10~GeV/$c$. Contributions involving pion decays in flight $\pi^\pm\to e^\pm\nu$ are found to be negligible. The standard $K^+\to\pi^+e^+e^-$ selection validates the modelling of backgrounds due to $\pi^\pm\to e^\pm$ misidentification with the LKr-based condition to 3\% precision. The control selection validates the modelling of backgrounds due to $\pi^+\to e^+$ misidentification with the RICH-based condition within the statistical precision. 

The number of $K^+$ decays in the FV is computed as
\begin{displaymath}
N_K = \frac{(1-f)\cdot N_{\pi ee}}{{\cal B}_{\pi ee}\cdot A_{\pi ee}} = (1.015 \pm 0.010_{\rm stat} \pm 0.030_{\rm syst})\times 10^{12},
\end{displaymath}
where $N_{\pi ee}=11041$ is the number of data $K^+\to\pi^+e^+e^-$ candidates in the signal $m_{\pi ee}$ region (constituting the world's largest sample of these decays), ${\cal B}_{\pi ee}=(3.00\pm0.09)\times 10^{-7}$ is the $K^+\to\pi^+e^+e^-$ branching fraction~\cite{pdg}, $A_{\pi ee}=(3.62\pm0.02_{\rm syst})\%$ is the selection acceptance evaluated with simulations including trigger inefficiency and random veto effects (with the uncertainty estimated by stability checks with respect to variation of the selection criteria), and $f=1.0\times 10^{-3}$ is the relative background contamination evaluated with simulations. The systematic error on $N_K$ is dominated by the external uncertainty on ${\cal B}_{\pi ee}$.
%

As a cross-check, the quantity $N_K$ is evaluated using the control selection including RICH-based $e^+$ identification. In this case, a 9\% fractional reduction of the acceptance $A_{\pi ee}$ is observed (mainly due to the track angular separation requirement in the RICH), $f$ is negligible, $N_{\pi ee}$ becomes 9922, and the resulting $N_K$ value changes by less than 1\%.


\boldmath
\section{Search for the $K^+\to\pi^-e^+e^+$ decay}
\unboldmath
\label{sec:lnv-piee}

The reconstructed $m_{\pi ee}$ spectra of the data, simulated signal and backgrounds obtained using the $K^+\to\pi^-e^+e^+$ selection are displayed in Fig.~\ref{fig:piee-lnv}~(left). Identification of two positrons with both LKr-based and RICH-based conditions leads to a stronger reduction of the $K^+\to\pi^+\pi^+\pi^-$ and $K^+\to\pi^+\pi^-e^+\nu$ backgrounds than in the $K^+\to\pi^+e^+e^-$ case. This makes the $K^+\to\pi^0_D e^+\nu$ decay the largest background source in the lower, masked and upper $m_{\pi ee}$ regions. The $K^+\to e^+\nu e^+e^-$ decay represents another background source in the masked region.

The $K^+\to\pi^0_D e^+\nu$ and $K^+\to e^+\nu e^+e^-$ backgrounds enter by $e^-\to\pi^-$ misidentification with the LKr-based condition, and their description relies on the $P_{e\pi}$ measurement and modelling (Fig.~\ref{fig:pid}, right). To validate the model, a control $K^+\to\pi^-e^+e^+$ selection is used, obtained from the standard selection by replacing the $p_{\pi ee}$ condition with a missing momentum requirement $p_{\rm beam}-p_{\pi ee} > 10~{\rm GeV}/c$ and removing the $p^{\pi ee}_T$ condition (thus removing the possible $K^+\to\pi^-e^+e^+$ signal contribution). The data and simulated $m_{\pi ee}$ spectra obtained using the control selection, and the ratio of these spectra, are shown in Fig.~\ref{fig:piee-lnv}~(right). The sample is dominated by $K^+\to\pi^0_D e^+\nu$ decays. The variation of the data/MC ratio over the $m_{\pi ee}$ range validates the background description to a 20\% relative precision. The limited accuracy is attributed to the MC description of the LAV detector response for soft photons from $K^+\to\pi^0_D e^+\nu$ decays. A 20\% relative systematic uncertainty is assigned to the estimated backgrounds to the signal decay $K^+\to\pi^-e^+e^+$ arising from $e^-\to\pi^-$ misidentification. 

The background due to pileup is found to be negligible by studying the sidebands of the track time distributions. Backgrounds from $K^+\to\pi^+\pi^0_{DD}$ decays and multiple photon conversions are studied using four control selections involving vertices with total charge $q\ne1$ ($\pi^-e^-e^-$, $\pi^-e^+e^-$, $\pi^+e^-e^-$ and $\pi^+e^+e^+$), without any $m_{\pi ee}$ selection criteria. The first two selections involving a $\pi^-$ are similar to the signal selection, and yield no events from the data sample. The last two selections allow the backgrounds from $K^+\to\pi^+\pi^0_{DD}$ and multiple conversions to enter without misidentification, enhancing the background by at least a factor of $10^4$. These selections yield 676 and 326 data events (mainly with $m_{ee}<140$~MeV/$c^2$), which is consistent with the $K^+\to\pi^+\pi^0_{DD}$ contribution expected from simulations. Considering the above enhancement factor, the background to the $K^+\to\pi^-e^+e^+$ signal is clearly negligible.

The estimated background contributions in the lower, upper, masked and signal $m_{\pi ee}$ regions are listed in Table~\ref{tab:bkg}. The numbers of data and expected background events in the lower and upper regions are compared before opening the masked region, and found to be in agreement within statistical fluctuations. The background in the signal region is estimated to be 
\begin{displaymath}
N_B = 0.43\pm 0.09,
\end{displaymath}
where the uncertainty is dominated by the systematic contribution due to the accuracy of $P_{e\pi}$ modelling. Background suppression in the signal region relies on the LKr-based $\pi^-$ identification and LAV photon veto conditions. After the masked region is opened, one data event is observed in this region but outside the signal region (in agreement with the expected background), and no events are observed in the signal region.

\begin{table}[tb]
\caption{Numbers of estimated background events and observed data events obtained using the $K^+\to\pi^-e^+e^+$ selection in the lower, upper, masked and signal $m_{\pi ee}$ regions. The uncertainties are dominated by the 20\% systematic error on the estimated $K^+\to\pi^0_De^+\nu$ and $K^+\to e^+\nu e^+e^-$ backgrounds (fully correlated between the two contributions). MC statistical errors and external uncertainties due to the background branching fractions and kinematic distributions~\cite{pdg} are negligible. The masked and signal regions are opened for the data sample only after the validation of the background estimates.}
\begin{center}
\vspace{-5mm}
\begin{tabular}{lcccc}
\hline
Mode & Lower region & Upper region & Masked region & Signal region \\
\hline
$K^+\to\pi^+\pi^+\pi^-$ & 0.9 & -- & -- & -- \\
$K^+\to\pi^+\pi^-e^+\nu$ & 3.3 & -- & -- & -- \\
$K^+\to\pi^+\pi^0_D$ & -- & 0.02 & 0.01 & -- \\
$K^+\to\pi^0_D e^+\nu$ & $3.7\pm0.7$ & $1.20\pm0.24$ & $1.23\pm0.25$ & $0.29\pm0.06$ \\
$K^+\to e^+\nu e^+e^-$ & $0.7\pm0.1$ & $0.76\pm0.15$ & $0.47\pm0.09$ & $0.14\pm0.03$ \\
\hline
Total & $8.6\pm0.9$ & $1.98\pm0.39$ & $1.71\pm0.34$ & $0.43\pm0.09$ \\
\hline
Data & 8 & 1 & 1 & 0 \\
\hline
\end{tabular}
\end{center}
\vspace{-5mm}
\label{tab:bkg}
\end{table}

The signal acceptance evaluated with simulations assuming a uniform phase space distribution is $A_{\pi ee}^{\rm LNV} = 4.32\%$, which is larger than $A_{\pi ee}$ due to the absence of the $m_{\pi ee}>140$~MeV/$c^2$ condition in the $K^+\to\pi^-e^+e^+$ selection. The single event sensitivity defined as the signal branching fraction corresponding to the observation of one signal event is found to be
\begin{displaymath}
{\cal B}_{\rm SES} = \frac{1}{N_K\cdot A_{\pi ee}^{\rm LNV}} = (2.28\pm0.07)\times 10^{-11}.
\end{displaymath}
An upper limit on the signal branching fraction is evaluated using the quantity ${\cal B}_{\rm SES}$ and the numbers of expected background events and observed data events using the CL$_{\rm S}$ method~\cite{re02}:
\begin{displaymath}
{\cal B}(K^+\to\pi^-e^+e^+) < 5.3 \times 10^{-11} ~~ {\rm at} ~ 90\% ~ {\rm CL}.
\end{displaymath}


\newpage

\boldmath
\section{Search for the $K^+\to\pi^-\pi^0e^+e^+$ decay}
\unboldmath
\label{sec:pipiee}

\begin{figure}[tb]
\begin{center}
\resizebox{0.5\textwidth}{!}{\includegraphics{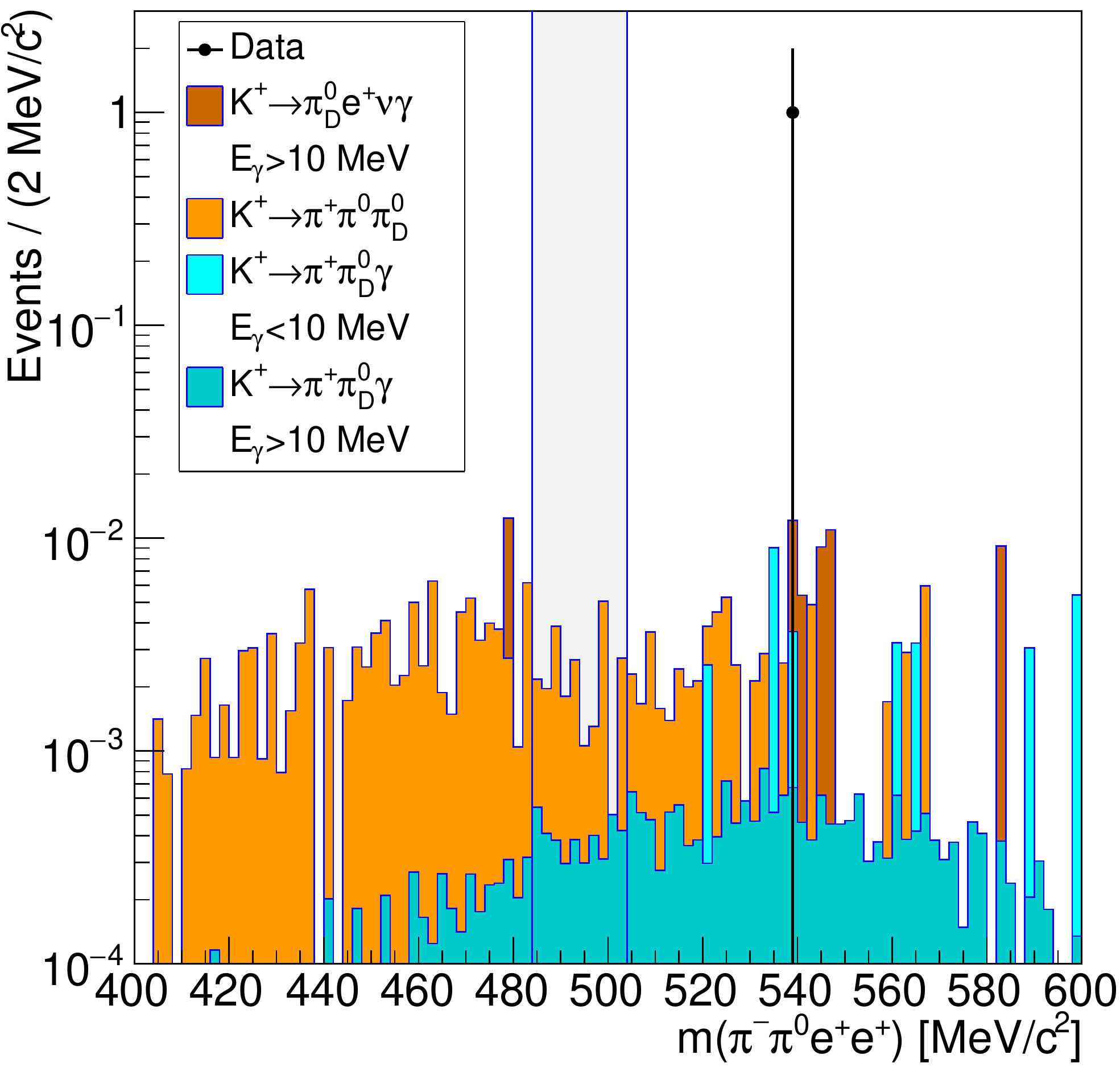}}%
\end{center}
\vspace{-14mm}
\caption{Reconstructed $m_{\pi\pi ee}$ spectra of the data and simulated backgrounds obtained using the $K^+\to\pi^-\pi^0e^+e^+$ selection. The $K^+\to\pi^+\pi^0 e^+e^-$ contribution is the smallest, and is not shown. The shaded vertical band indicates the signal $m_{\pi\pi ee}$ region masked during the analysis.}
\label{fig:pipiee-lnv}
\end{figure}

The reconstructed $m_{\pi\pi ee}$ spectra of the data, simulated signal and backgrounds  obtained using the $K^+\to\pi^-\pi^0e^+e^+$ selection are shown in Fig.~\ref{fig:pipiee-lnv}. Background sources studied with simulations include $K^+\to\pi^+\pi^0\pi^0_D$, $K^+\to\pi^+\pi^0_D\gamma$ and $K^+\to\pi^+\pi^0 e^+e^-$ decays with double particle misidentification ($\pi^+\to e^+$ and $e^-\to\pi^-$), and the $K^+\to\pi^0_D e^+\nu\gamma$ decay with single misidentification ($e^-\to\pi^-$). The two background sources with a radiative photon (which gives rise to a photon candidate in $\pi^0\to\gamma\gamma$ reconstruction) are dominated by the inner bremsstrahlung components. To minimise MC statistical errors, $K^+\to\pi^+\pi^0_D\gamma$ components with the radiative photon energy in the $K^+$ rest frame, $E_\gamma$, below and above 10~MeV are simulated separately; the two contributions are found to be of similar size. The $K^+\to\pi^0_D e^+\nu\gamma$ contribution entering via single misidentification ($e^-\to\pi^-$) is evaluated for $E_\gamma>10$~MeV only due to computational limitations; all simulated events passing the selection are found to have $E_\gamma>25$~MeV.

\begin{table}[tb]
\caption{Numbers of estimated background events and observed data events obtained using the $K^+\to\pi^-\pi^0e^+e^+$ selection in the control and signal $m_{\pi\pi ee}$ regions. The signal region is unmasked for the data sample only after the validation of the background estimate.}
\begin{center}
\vspace{-7mm}
\begin{tabular}{lcc}
\hline
Mode & Control region & Signal region \\
\hline
$K^+\to\pi^+\pi^0\pi^0_D$ & $0.16\pm0.01$ & 0.019 \\
$K^+\to\pi^+\pi^0_D\gamma$ & $0.06\pm0.01$ & 0.004 \\
$K^+\to\pi^0_D e^+\nu\gamma$ & $0.05\pm0.02$ & -- \\
$K^+\to\pi^+\pi^0 e^+e^-$ & 0.01 & 0.001 \\
Pileup & $0.20\pm0.20$ & $0.020\pm0.020$ \\
\hline
Total & $0.48\pm0.20$ & $0.044\pm0.020$ \\
\hline
Data & 1 & 0 \\
\hline
\end{tabular}
\end{center}
\vspace{-7mm}
\label{tab:bkg2}
\end{table}

Background due to pileup is potentially larger than in the $K^+\to\pi^-e^+e^+$ case due to the absence of topological constraints on the photon candidates, as opposed to the three-track vertex condition for the tracks. This background is studied with the data, using two control selections with inverted photon timing conditions. In the first (second) of these selections, one photon is (both photons are) required to be in the sidebands of the photon time distribution, separated by 5--30~ns from the vertex time. The selection with one out-of-time photon yields one data event in the control mass region at $m_{\pi\pi ee}=573$~MeV/$c^2$ and no events in the signal region, while the selection with both photons out of time yields no data events. Background contamination due to pileup is estimated considering that sidebands are five times wider than the signal time window, and assuming that the background $m_{\pi\pi ee}$ distribution is uniform over the control and signal regions.

The estimated background contributions in the control and signal regions are summarised in Table~\ref{tab:bkg2}. One data event is observed in the control region, in agreement with the background expectation. The total expected background in the signal region is found to be 
\begin{displaymath}
N_B = 0.044\pm 0.020,
\end{displaymath}
where the uncertainty is dominated by the statistical error in the pileup background estimate. After unmasking, no data events are found in the signal region.

The signal acceptance evaluated with simulation assuming a uniform phase space distribution is $A_{\pi\pi ee}^{\rm LNV} = (0.271\pm0.003_{\rm syst})\%$. The uncertainty is due to the $\pi^0\to\gamma\gamma$ decay reconstruction (which does not cancel between signal and normalisation), and is evaluated from a study of $K^+\to\pi^0e^+\nu$ decays in which variations to LKr calorimeter simulated response and calibration are included. The acceptance is suppressed with respect to the decays without $\pi^0$ emission ($A_{\pi\pi ee}^{\rm LNV}/A_{\pi ee}^{\rm LNV}\approx 0.06$) by the geometric acceptance for the photons from $\pi^0\to\gamma\gamma$ decay. Note that the acceptance for the SM decay $K^+\to\pi^+\pi^0e^+e^-$ achieved with a selection similar to the signal one, i.e. based on identification of each track, is ${\cal O}(10^{-5})$ due to the predominantly soft $e^+e^-$ pairs. Therefore the $K^+\to\pi^+\pi^0e^+e^-$ decay is not used for normalisation.

The single event sensitivity is evaluated as
\begin{displaymath}
{\cal B}_{\rm SES} = \frac{1}{N_K\cdot {\cal B}_{\gamma\gamma} \cdot A_{\pi\pi ee}^{\rm LNV}} = (3.68\pm0.12)\times 10^{-10},
\end{displaymath}
where ${\cal B}_{\gamma\gamma}=(98.823 \pm 0.034)\%$ is the $\pi^0\to\gamma\gamma$ branching fraction~\cite{pdg}. An upper limit on the signal branching fraction is evaluated using the CL$_{\rm S}$ method:
\begin{displaymath}
{\cal B}(K^+\to\pi^-\pi^0e^+e^+) < 8.5 \times 10^{-10} ~~ {\rm at} ~ 90\% ~ {\rm CL}.
\end{displaymath}
For comparison, the branching fraction of the corresponding SM decay is ${\cal B}(K^+\to\pi^+\pi^0e^+e^-)=(4.24\pm0.14)\times 10^{-6}$, as measured by the NA48/2 experiment~\cite{ba19}.


\section*{Summary}

Searches for lepton number violating $K^+\to\pi^-e^+e^+$ and $K^+\to\pi^-\pi^0e^+e^+$ decays have been performed using the complete dataset collected by the NA62 experiment at CERN in 2016--2018. Upper limits of $5.3\times 10^{-11}$ and $8.5\times 10^{-10}$ were obtained on the decay branching fractions at 90\% confidence level, assuming uniform phase space distributions. The former result improves by a factor of four over the previous best limit~\cite{co19}, while the latter result represents the first limit on the $K^+\to\pi^-\pi^0e^+e^+$ decay rate. The sensitivity of both searches is limited by the size of the dataset and not by backgrounds. Similarly to most other limits for LNV decay rates, these results depend on phase space density assumptions.


\section*{Acknowledgements}

\input{acknow202108}




\end{document}

%% file: authors.tex
\begin{center}
{\Large The NA62 Collaboration$\,$\renewcommand{\thefootnote}{\fnsymbol{footnote}}%
\footnotemark[1]\renewcommand{\thefootnote}{\arabic{footnote}}}\\
\end{center}


\begin{raggedright}
\noindent
{\bf Universit\'e Catholique de Louvain, Louvain-La-Neuve, Belgium}\\
 \mbox{E.~Cortina Gil,
 A.~Kleimenova,
 E.~Minucci$\,$\footnotemark[1]$^,\,$\footnotemark[2],
 S.~Padolski$\,$\footnotemark[3],
 P.~Petrov,
 A.~Shaikhiev$\,$\footnotemark[4],
 R.~Volpe$\,$\footnotemark[5]}\\[2mm]

{\bf TRIUMF, Vancouver, British Columbia, Canada}\\
 T.~Numao,
 Y.~Petrov,
 B.~Velghe,
 V. W. S.~Wong\\[2mm]

{\bf University of British Columbia, Vancouver, British Columbia, Canada}\\
 D.~Bryman$\,$\footnotemark[6],
 J.~Fu\\[2mm]

{\bf Charles University, Prague, Czech Republic}\\
 T.~Husek$\,$\footnotemark[7],
 J.~Jerhot$\,$\footnotemark[8],
 K.~Kampf,
 M.~Zamkovsky$\,$\footnotemark[8]\\[2mm]

{\bf Institut f\"ur Physik and PRISMA Cluster of Excellence, Universit\"at Mainz, Mainz, Germany}\\
 R.~Aliberti$\,$\footnotemark[9],
 G.~Khoriauli$\,$\footnotemark[10],
 J.~Kunze,
 D.~Lomidze$\,$\footnotemark[11],
 L.~Peruzzo,
 M.~Vormstein,
 R.~Wanke\\[2mm]

{\bf Dipartimento di Fisica e Scienze della Terra dell'Universit\`a e INFN, Sezione di Ferrara, Ferrara, Italy}\\
 P.~Dalpiaz,
 M.~Fiorini,
 I.~Neri,
 A.~Norton$\,$\footnotemark[12],
 F.~Petrucci,
 H.~Wahl$\,$\footnotemark[13]\\[2mm]

{\bf INFN, Sezione di Ferrara, Ferrara, Italy}\\
 A.~Cotta Ramusino,
 A.~Gianoli\\[2mm]

{\bf Dipartimento di Fisica e Astronomia dell'Universit\`a e INFN, Sezione di Firenze, Sesto Fiorentino, Italy}\\
 E.~Iacopini,
 G.~Latino,
 M.~Lenti,
 A.~Parenti\\[2mm]

{\bf INFN, Sezione di Firenze, Sesto Fiorentino, Italy}\\
 A.~Bizzeti$\,$\footnotemark[14],
 F.~Bucci\\[2mm]

{\bf Laboratori Nazionali di Frascati, Frascati, Italy}\\
 A.~Antonelli,
 G.~Georgiev$\,$\footnotemark[15],
 V.~Kozhuharov$\,$\footnotemark[15],
 G.~Lanfranchi,
 S.~Martellotti,
 M.~Moulson,
 T.~Spadaro,
 G.~Tinti\\[2mm]

{\bf Dipartimento di Fisica ``Ettore Pancini'' e INFN, Sezione di Napoli, Napoli, Italy}\\
 F.~Ambrosino,
 T.~Capussela,
 M.~Corvino$\,$\footnotemark[1],
 D.~Di Filippo,
 R.~Fiorenza$\,$\footnotemark[16],
 P.~Massarotti,
 M.~Mirra,
 M.~Napolitano,
 G.~Saracino\\[2mm]

{\bf Dipartimento di Fisica e Geologia dell'Universit\`a e INFN, Sezione di Perugia, Perugia, Italy}\\
 G.~Anzivino,
 F.~Brizioli,
 E.~Imbergamo,
 R.~Lollini,
 R.~Piandani$\,$\footnotemark[17],
 C.~Santoni\\[2mm]

{\bf INFN, Sezione di Perugia, Perugia, Italy}\\
 M.~Barbanera,
 P.~Cenci,
 B.~Checcucci,
 P.~Lubrano,
 M.~Lupi$\,$\footnotemark[18],
 M.~Pepe,
 M.~Piccini\\[2mm]

{\bf Dipartimento di Fisica dell'Universit\`a e INFN, Sezione di Pisa, Pisa, Italy}\\
 F.~Costantini,
 L.~Di Lella$\,$\footnotemark[13],
 N.~Doble$\,$\footnotemark[13],
 M.~Giorgi,
 S.~Giudici,
 G.~Lamanna,
 E.~Lari,
 E.~Pedreschi,
 M.~Sozzi\\[2mm]

{\bf INFN, Sezione di Pisa, Pisa, Italy}\\
 C.~Cerri,
 R.~Fantechi,
 L.~Pontisso$\,$\footnotemark[19],
 F.~Spinella\\[2mm]

{\bf Scuola Normale Superiore e INFN, Sezione di Pisa, Pisa, Italy}\\
 I.~Mannelli\\[2mm]

\newpage

{\bf Dipartimento di Fisica, Sapienza Universit\`a di Roma e INFN, Sezione di Roma I, Roma, Italy}\\
 G.~D'Agostini,
 M.~Raggi\\[2mm]

{\bf INFN, Sezione di Roma I, Roma, Italy}\\
 A.~Biagioni,
 P.~Cretaro,
 O.~Frezza,
 E.~Leonardi,
 A.~Lonardo,
 M.~Turisini,
 P.~Valente,
 P.~Vicini\\[2mm]

{\bf INFN, Sezione di Roma Tor Vergata, Roma, Italy}\\
 R.~Ammendola,
 V.~Bonaiuto$\,$\footnotemark[20],
 A.~Fucci,
 A.~Salamon,
 F.~Sargeni$\,$\footnotemark[21]\\[2mm]

{\bf Dipartimento di Fisica dell'Universit\`a e INFN, Sezione di Torino, Torino, Italy}\\
 R.~Arcidiacono$\,$\footnotemark[22],
 B.~Bloch-Devaux,
 M.~Boretto$\,$\footnotemark[1],
 E.~Menichetti,
 E.~Migliore,
 D.~Soldi\\[2mm]

{\bf INFN, Sezione di Torino, Torino, Italy}\\
 C.~Biino,
 A.~Filippi,
 F.~Marchetto\\[2mm]

{\bf \mbox{Instituto de F\'isica, Universidad Aut\'onoma de San Luis Potos\'i, San Luis Potos\'i, Mexico}}\\
 J.~Engelfried,
 N.~Estrada-Tristan$\,$\footnotemark[23]\\[2mm]

{\bf Horia Hulubei National Institute of Physics for R\&D in Physics and Nuclear Engineering, Bucharest-Magurele, Romania}\\
 A. M.~Bragadireanu,
 S. A.~Ghinescu,
 O. E.~Hutanu\\[2mm]

{\bf Joint Institute for Nuclear Research, Dubna, Russia}\\
 A.~Baeva,
 D.~Baigarashev$\,$\footnotemark[24],
 D.~Emelyanov,
 T.~Enik,
 V.~Falaleev$\,$\footnotemark[25],
 V.~Kekelidze,
 A.~Korotkova,
 L.~Litov$\,$\footnotemark[15],
 D.~Madigozhin,
 M.~Misheva$\,$\footnotemark[26],
 N.~Molokanova,
 S.~Movchan,
 I.~Polenkevich,
 Yu.~Potrebenikov,
 S.~Shkarovskiy,
 A.~Zinchenko$\,$\renewcommand{\thefootnote}{\fnsymbol{footnote}}\footnotemark[2]\renewcommand{\thefootnote}{\arabic{footnote}}\\[2mm]

{\bf \mbox{Institute for Nuclear Research of the Russian Academy of Sciences, Moscow, Russia}}\\
 \mbox{S.~Fedotov,
 E.~Gushchin,
 A.~Khotyantsev,
 Y.~Kudenko$\,$\footnotemark[27],
 V.~Kurochka,
 M.~Medvedeva,
 A.~Mefodev}\\[2mm]

{\bf Institute for High Energy Physics of the Russian Federation State Research Center ``Kurchatov Institute", Protvino, Russia}\\
 S.~Kholodenko,
 V.~Kurshetsov,
 V.~Obraztsov,
 A.~Ostankov$\,$\renewcommand{\thefootnote}{\fnsymbol{footnote}}\footnotemark[2]\renewcommand{\thefootnote}{\arabic{footnote}},
 V.~Semenov$\,$\renewcommand{\thefootnote}{\fnsymbol{footnote}}\footnotemark[2]\renewcommand{\thefootnote}{\arabic{footnote}},
 V.~Sugonyaev,
 O.~Yushchenko\\[2mm]

{\bf Faculty of Mathematics, Physics and Informatics, Comenius University, Bratislava, Slovakia}\\
 L.~Bician$\,$\footnotemark[1],
 T.~Blazek,
 V.~Cerny,
 Z.~Kucerova\\[2mm]

{\bf CERN,  European Organization for Nuclear Research, Geneva, Switzerland}\\
 J.~Bernhard,
 A.~Ceccucci,
 H.~Danielsson,
 N.~De Simone$\,$\footnotemark[28],
 F.~Duval,
 B.~D\"obrich,
 L.~Federici,
 E.~Gamberini,
 L.~Gatignon$\,$\footnotemark[29],
 R.~Guida,
 F.~Hahn$\,$\renewcommand{\thefootnote}{\fnsymbol{footnote}}\footnotemark[2]\renewcommand{\thefootnote}{\arabic{footnote}},
 E. B.~Holzer,
 B.~Jenninger,
 M.~Koval$\,$\footnotemark[30],
 P.~Laycock$\,$\footnotemark[3],
 G.~Lehmann Miotto,
 P.~Lichard,
 A.~Mapelli,
 R.~Marchevski$\,$\footnotemark[31],
 K.~Massri,
 M.~Noy,
 V.~Palladino$\,$\footnotemark[32],
 M.~Perrin-Terrin$\,$\footnotemark[33]$^,\,$\footnotemark[34],
 J.~Pinzino$\,$\footnotemark[35],
 V.~Ryjov,
 S.~Schuchmann$\,$\footnotemark[13],
 S.~Venditti\\[2mm]

{\bf University of Birmingham, Birmingham, United Kingdom}\\
 T.~Bache,
 M. B.~Brunetti$\,$\footnotemark[36],
 V.~Duk$\,$\footnotemark[37],
 V.~Fascianelli$\,$\footnotemark[38],
 J. R.~Fry,
 F.~Gonnella,
 E.~Goudzovski$\renewcommand{\thefootnote}{\fnsymbol{footnote}}\footnotemark[1]\renewcommand{\thefootnote}{\arabic{footnote}}$,
 J.~Henshaw,
 L.~Iacobuzio,
 C.~Lazzeroni,
 N.~Lurkin$\,$\footnotemark[8],
 F.~Newson,
 C.~Parkinson,
 A.~Romano,
 A.~Sergi$\,$\footnotemark[39],
 A.~Sturgess,
 J.~Swallow$\,$\footnotemark[1],
 A.~Tomczak\\[2mm]

{\bf University of Bristol, Bristol, United Kingdom}\\
 H.~Heath,
 R.~Page,
 S.~Trilov\\[2mm]

{\bf University of Glasgow, Glasgow, United Kingdom}\\
 B.~Angelucci,
 D.~Britton,
 C.~Graham,
 D.~Protopopescu\\[2mm]

\newpage
{\bf University of Lancaster, Lancaster, United Kingdom}\\
 J.~Carmignani,
 J. B.~Dainton,
 R. W. L.~Jones,
 G.~Ruggiero$\,$\footnotemark[31]\\[2mm]

{\bf University of Liverpool, Liverpool, United Kingdom}\\
 L.~Fulton,
 D.~Hutchcroft,
 E.~Maurice$\,$\footnotemark[40],
 B.~Wrona\\[2mm]

{\bf George Mason University, Fairfax, Virginia, USA}\\
 A.~Conovaloff,
 P.~Cooper,
 D.~Coward$\,$\footnotemark[41],
 P.~Rubin\\[2mm]

\end{raggedright}
%
%
\setcounter{footnote}{0}
\renewcommand{\thefootnote}{\fnsymbol{footnote}}
\footnotetext[1]{Corresponding author: E.~Goudzovski, email: evgueni.goudzovski@cern.ch}
\footnotetext[2]{Deceased}
\renewcommand{\thefootnote}{\arabic{footnote}}
\footnotetext[1]{Present address: CERN,  European Organization for Nuclear Research, CH-1211 Geneva 23, Switzerland}
\footnotetext[2]{Also at Laboratori Nazionali di Frascati, I-00044 Frascati, Italy}
\footnotetext[3]{Present address: Brookhaven National Laboratory, Upton, NY 11973, USA}
\footnotetext[4]{Present address: School of Physics and Astronomy, University of Birmingham, Birmingham, B15 2TT, UK}
\footnotetext[5]{Present address: Faculty of Mathematics, Physics and Informatics, Comenius University, 842 48, Bratislava, Slovakia}
\footnotetext[6]{Also at TRIUMF, Vancouver, British Columbia, V6T 2A3, Canada}
\footnotetext[7]{Present address: Department of Astronomy and Theoretical Physics, Lund University, Lund, SE 223-62, Sweden}
\footnotetext[8]{Present address: Universit\'e Catholique de Louvain, B-1348 Louvain-La-Neuve, Belgium}
\footnotetext[9]{Present address: Institut f\"ur Kernphysik and Helmholtz Institute Mainz, Universit\"at Mainz, Mainz, D-55099, Germany}
\footnotetext[10]{Present address: Universit\"at W\"urzburg, D-97070 W\"urzburg, Germany}
\footnotetext[11]{Present address: European XFEL GmbH, D-22761 Hamburg, Germany}
\footnotetext[12]{Present address: University of Glasgow, Glasgow, G12 8QQ, UK}
\footnotetext[13]{Present address: Institut f\"ur Physik and PRISMA Cluster of Excellence, Universit\"at Mainz, D-55099 Mainz, Germany}
\footnotetext[14]{Also at Dipartimento di Fisica, Universit\`a di Modena e Reggio Emilia, I-41125 Modena, Italy}
\footnotetext[15]{Also at Faculty of Physics, University of Sofia, BG-1164 Sofia, Bulgaria}
\footnotetext[16]{Present address: Scuola Superiore Meridionale e INFN, Sezione di Napoli, I-80138 Napoli, Italy}
\footnotetext[17]{Present address: University of Chinese Academy of Sciences, Beijing, 100049, China}
\footnotetext[18]{Present address: Institut am Fachbereich Informatik und Mathematik, Goethe Universit\"at, D-60323 Frankfurt am Main, Germany}
\footnotetext[19]{Present address: INFN, Sezione di Roma I, I-00185 Roma, Italy}
\footnotetext[20]{Also at Department of Industrial Engineering, University of Roma Tor Vergata, I-00173 Roma, Italy}
\footnotetext[21]{Also at Department of Electronic Engineering, University of Roma Tor Vergata, I-00173 Roma, Italy}
\footnotetext[22]{Also at Universit\`a degli Studi del Piemonte Orientale, I-13100 Vercelli, Italy}
\footnotetext[23]{Also at Universidad de Guanajuato, 36000 Guanajuato, Mexico}
\footnotetext[24]{Also at L.N. Gumilyov Eurasian National University, 010000 Nur-Sultan, Kazakhstan}
\footnotetext[25]{Also at Institute for Nuclear Research of the Russian Academy of Sciences, 117312 Moscow, Russia}
\footnotetext[26]{Present address: Institute of Nuclear Research and Nuclear Energy of Bulgarian Academy of Science (INRNE-BAS), BG-1784 Sofia, Bulgaria}
\footnotetext[27]{Also at National Research Nuclear University (MEPhI), 115409 Moscow and Moscow Institute of Physics and Technology, 141701 Moscow region, Moscow, Russia}
\footnotetext[28]{Present address: DESY, D-15738 Zeuthen, Germany}
\footnotetext[29]{Present address: University of Lancaster, Lancaster, LA1 4YW, UK}
\footnotetext[30]{Present address: Charles University, 116 36 Prague 1, Czech Republic}
\footnotetext[31]{Present address: Dipartimento di Fisica e Astronomia dell'Universit\`a e INFN, Sezione di Firenze, I-50019 Sesto Fiorentino, Italy}
\footnotetext[32]{Present address: Physics Department, Imperial College London, London, SW7 2BW, UK}
\footnotetext[33]{Present address: Aix Marseille University, CNRS/IN2P3, CPPM, F-13288, Marseille, France}
\footnotetext[34]{Also at Universit\'e Catholique de Louvain, B-1348 Louvain-La-Neuve, Belgium}
\footnotetext[35]{Present address: INFN, Sezione di Pisa, I-56100 Pisa, Italy}
\footnotetext[36]{Present address: Department of Physics, University of Warwick, Coventry, CV4 7AL, UK}
\footnotetext[37]{Present address: INFN, Sezione di Perugia, I-06100 Perugia, Italy}
\footnotetext[38]{Present address: Center for theoretical neuroscience, Columbia University, New York, NY 10027, USA}
\footnotetext[39]{Present address: Dipartimento di Fisica dell'Universit\`a e INFN, Sezione di Genova, I-16146 Genova, Italy}
\footnotetext[40]{Present address: Laboratoire Leprince Ringuet, F-91120 Palaiseau, France}
\footnotetext[41]{Also at SLAC National Accelerator Laboratory, Stanford University, Menlo Park, CA 94025, USA}

%% file: acknow202108.tex
It is a pleasure to express our appreciation to the staff of the CERN laboratory and the technical
staff of the participating laboratories and universities for their efforts in the operation of the
experiment and data processing.

The cost of the experiment and its auxiliary systems was supported by the funding agencies of 
the Collaboration Institutes. We are particularly indebted to: 
F.R.S.-FNRS (Fonds de la Recherche Scientifique - FNRS), under Grants No. 4.4512.10, 1.B.258.20, Belgium;
CECI (Consortium des Equipements de Calcul Intensif), funded by the Fonds de la Recherche Scientifique de Belgique (F.R.S.-FNRS) under Grant No. 2.5020.11 and by the Walloon Region, Belgium;
NSERC (Natural Sciences and Engineering Research Council), funding SAPPJ-2018-0017,  Canada;
MEYS (Ministry of Education, Youth and Sports) funding LM 2018104, Czech Republic;
BMBF (Bundesministerium f\"{u}r Bildung und Forschung) contracts 05H12UM5, 05H15UMCNA and 05H18UMCNA, Germany;
INFN  (Istituto Nazionale di Fisica Nucleare),  Italy;
MIUR (Ministero dell'Istruzione, dell'Universit\`a e della Ricerca),  Italy;
CONACyT  (Consejo Nacional de Ciencia y Tecnolog\'{i}a),  Mexico;
IFA (Institute of Atomic Physics) Romanian CERN-RO No. 1/16.03.2016 and Nucleus Programme PN 19 06 01 04,  Romania;
INR-RAS (Institute for Nuclear Research of the Russian Academy of Sciences), Moscow, Russia; 
JINR (Joint Institute for Nuclear Research), Dubna, Russia; 
NRC (National Research Center)  ``Kurchatov Institute'' and MESRF (Ministry of Education and Science of the Russian Federation), Russia; 
MESRS  (Ministry of Education, Science, Research and Sport), Slovakia; 
CERN (European Organization for Nuclear Research), Switzerland; 
STFC (Science and Technology Facilities Council), United Kingdom;
NSF (National Science Foundation) Award Numbers 1506088 and 1806430,  U.S.A.;
ERC (European Research Council)  ``UniversaLepto'' advanced grant 268062, ``KaonLepton'' starting grant 336581, Europe.

Individuals have received support from:
Charles University Research Center (UNCE/SCI/013), Czech Republic;
Ministero dell'Istruzione, dell'Universit\`a e della Ricerca (MIUR  ``Futuro in ricerca 2012''  grant RBFR12JF2Z, Project GAP), Italy;
Russian Science Foundation (RSF 19-72-10096), Russia;
the Royal Society  (grants UF100308, UF0758946), United Kingdom;
STFC (Rutherford fellowships ST/J00412X/1, ST/M005798/1), United Kingdom;
ERC (grants 268062,  336581 and  starting grant 802836 ``AxScale'');
EU Horizon 2020 (Marie Sk\l{}odowska-Curie grants 701386, 754496, 842407, 893101, 101023808).